\documentclass[journal]{IEEEtran}

\ifCLASSINFOpdf
\else
   \usepackage[dvips]{graphicx}
\fi

\usepackage{url}
\usepackage{graphicx}
\usepackage{amsmath}
\usepackage[ruled,vlined]{algorithm2e}
\usepackage{booktabs}
\newcommand{\vect}[1]{\mathbf{#1}}
\newcommand{\inv}{^{\text{-}1}}

\begin{document}

\title{Xi-Vector Embedding for Speaker Recognition}

\author{Kong Aik Lee, \IEEEmembership{Senior Member, IEEE}, Qiongqiong Wang, and Takafumi Koshinaka
\thanks{K. A. Lee was with the Biometrics Research Laboratories, NEC Corporation, Japan. He is now with the Institute for Infocomm Research, A*STAR, Singapore (e-mail: lee\_kong\_aik@i2r.a-star.edu.sg).}
\thanks{Q. Wang is with the Biometrics Research Laboratories, NEC Corporation, Japan (e-mail: q-wang@nec.com).}
\thanks{K. Takafumi was with the Biometrics Research Laboratories, NEC Corporation, Japan. He is now with the School of Data Science, Yokohama City University, Japan (e-mail: koshinak@yokohama-cu.ac.jp).}}
\markboth{IEEE SPL, June. 2021}
{Shell \MakeLowercase{\textit{et al.}}: Xi-Vector Embedding for Speaker Verification}
\maketitle

\begin{abstract}
We present a Bayesian formulation for deep speaker embedding, wherein the xi-vector is the Bayesian counterpart of the x-vector, taking into account the uncertainty estimate. On the technology front, we offer a simple and straightforward extension to the now widely used x-vector. It consists of an auxiliary neural net predicting the frame-wise uncertainty of the input sequence. We show that the proposed extension leads to substantial improvement across all operating points, with a significant reduction in error rates and detection cost. On the theoretical front, our proposal integrates the Bayesian formulation of linear Gaussian model to speaker-embedding neural networks via the pooling layer. In one sense, our proposal integrates the Bayesian formulation of the i-vector to that of the x-vector. Hence, we refer to the embedding as the xi-vector, which is pronounced as /zai/ vector. Experimental results on the SITW evaluation set show a consistent improvement of over $17.5\%$ in equal-error-rate and $10.9\%$ in minimum detection cost.     
\end{abstract}


\IEEEpeerreviewmaketitle

\section{Introduction}

Automatic speaker recognition is the task of identifying or verifying an individual’s identity from samples of his/her voice using machine learning algorithms, without any human intervention~\cite{lee2020a}. How speaker recognition is carried out has changed substantially throughout these years with increasing accuracy and robustness against various sources of variability in the speech signal. Modern speaker recognition systems \cite{i4u2019,Matejka2020,Villalba2020,Nagrani2020,lee2020nec} consist of a speaker embedding front-end followed by a scoring backend. In this approach, speech utterances are first represented as fixed-length vectors -- the so-called speaker embeddings. The current de-facto standard of speaker embedding is x-vector~\cite{snyder2018vector}. For scoring backend, probabilistic linear discrimination analysis (PLDA)~\cite{ioffe2006plda,Princepaper} is commonly used.

Unlike passwords that have zero uncertainty conditioned on a person’s identity~\cite{TAKAHASHI2014}, human voices exhibit both extrinsic and intrinsic variability. Major sources of extrinsic variability in speech signals are background noises, channel distortion, and room acoustics. Intrinsic factors include the physiological nature of the vocal apparatus and psychological states (e.g., emotion, mental health condition, etc.) of the speaker, and the biological constraint of the vocal tract leads to acoustically different utterances every time we repeat the same sentence. X-vectors, and similar forms of deep speaker embedding, do not consider the uncertainty of features. In a restricted sense, uncertainty is merely captured implicitly with empirical variance estimates at the utterance level. Consequently, they show low robustness against local and random perturbation which is the inherent property of speech utterances. 

The ability to handle uncertainty has been the cornerstone in the successful use of generative models for speaker recognition \cite{Reynolds00,kenny2008study,Dehak10frontend}. In the \emph{universal background model} or UBM, the uncertainty of feature vectors is modeled with the covariance matrices associated with its Gaussian components~\cite{Reynolds00}. In the i-vector embedding, uncertainty is captured with the prior and posterior covariance matrices of its latent variable \cite{kenny2008study,Dehak10frontend}.

Under the big data regime, deep learning is used to achieve state-of-the-art performance, notwithstanding that most speaker-embedding neural networks are not able to represent uncertainty. In this paper, we bridge the gap by incorporating uncertainty modeling in a speaker embedding neural network. To this end, we first estimate the frame-wise uncertainty of the input sequences with an auxiliary neural network. This operation is handled with a linear Gaussian model at the pooling layer and trained as part of the neural network.


In addition to good performance, our study sheds light on the role of uncertainty in speaker embedding. It turns out that the variance estimate is used as the indicator to which feature vectors, and which dimensions of the feature vectors, are useful in speaker classification. Higher weights are given to those frames and features with lower uncertainty and vice versa. In a way, it plays a role similar to the attention model reported in~\cite{okabe2018attentive,qwang2018}. The difference here is that the usefulness, or importance of feature vectors, is associated with the uncertainty estimate while this is loosely defined in prior works. Furthermore, the prior in the linear Gaussian model has its effect on the weights assigned to each frame and feature. More importantly, we show that a generative model could be inserted as part of a discriminative neural network to handle uncertainty.

\vspace{-2ex}
\section{Neural Speaker Embeddings}

We define speaker embedding vectors as the representation of variable-length utterances as fixed-length continuous-valued vectors. Embeddings from the same speakers are close together in the embedding vector space, and therefore allow easy comparison between speakers with simple geometric operations. Depending on how the embedding extractor is trained, we divide speaker embeddings into two broad categories, namely, (i) unsupervised embedding, and (ii) supervised embedding. The celebrated i-vector~\cite{Dehak10frontend}, and the classical \emph{Gaussian mixture model} (GMM) supervector~\cite{campbell2006} and alike~\cite{lee2011}, belong to the first category. A popular example of the second category is the x-vector embeddings~\cite{snyder2018vector, Desplanques2020}. Different from that of the latter, both i-vector and GMM supervector are based on a generative model trained using an unsupervised maximum likelihood criterion. Supervised embeddings rely on the use of labelled data for discriminative training, typically, with a multi-class speaker-discriminative loss. 


Recent development has shown the benefit of supervised learning with deep neural networks (DNNs) coupled with the massive use of data augmentation. Both have advanced the performance of neural speaker embedding by a large margin. The conventional x-vector extractor is a DNN consisting of three functional blocks: 
\begin{itemize}
    \item An \textbf{encoder} (or a frame processor) implemented with multiple layers of time-delay neural network (TDNN) \cite{Waibel89,tdnn}. In \cite{tcn}, it's shown that a TDNN could be implemented as a 1D-CNN with dilation. In~\cite{Villalba2020, zeinali2019, Chung2018}, 2D-convolution has shown to be effective as well.    
    \item A \textbf{temporal pooling layer} to compute an aggregated measure from the frame-level feature vectors produced by the encoder.
    \item A \textbf{decoder} to classify the utterances to speaker classes. One of the layers is designed to be a bottleneck layer, the output of which (after affine projection, and before the non-linearity) is the so-called x-vector speaker embedding. 
\end{itemize}
One element that is missing in the current framework is the ability to handle uncertainty induced by the random factors inherent in the generation (intrinsic) and transmission (extrinsic) of human voices. In particular, the frame-level features produced by the encoder are point estimates, which do not consider the uncertainty of the latent representation. In this paper, we bridge the gap with the provision of (i.) an encoder that infers the uncertainty in addition to the point estimate of frame-level features, and (ii.) a generative model that leverages the uncertainty measure in deep speaker embedding. It is worth mentioning that the frame-wise uncertainty estimated by the encoder neural network is regarded as \emph{aleatoric} uncertainty~\cite{Kendall2017} in Bayesian deep learning.



\section{Uncertainty modeling in the latent space}

To incorporate uncertainty measurement and modeling into the neural speaker embedding framework, we introduce two new concepts here.  

\subsection{Uncertainty estimation}
The encoder maps an input frame $\vect{x}_t$ to a point estimate $\vect{z}_t$ in a latent space. We propose to characterize the frame uncertainty with a covariance matrix $\vect{L}\inv_t$ associated with each estimate $\vect{z}_t$. This is accomplished by using a base neural network with two heads, one for $\vect{z}_t$ and one for $\vect{L}\inv_t$, and train simultaneously. Without loss of generality, we denote these operations as
\begin{equation}
\begin{split}
   \vect{z}_t = f_{\mathrm{enc}}(\vect{x}_t|\vect{x}_{t}^{t\pm n}) \quad\mathrm{and}\quad 
   \log\,\vect{L}_t = g_{\mathrm{enc}}(\vect{x}_t|\vect{x}_{t}^{t\pm n})    
\end{split}    
\label{eq:frameprocessor}
\end{equation}
where a context of $\pm n$ neighbouring frames is taken into account for each estimate. Note that the second equation in \eqref{eq:frameprocessor} is absent in the conventional x-vector embedding. Also, we assume that the covariance (and precision) matrices are diagonal and chose to estimate directly the log-precision which turns out to be more convenient as we shall see in the next section. 

With the construct in~\eqref{eq:frameprocessor}, the frame encoder maps an input sequence $\{\vect{x}_1, \vect{x}_2,\allowbreak ...\vect{x}_T\}$ of length $T$ to another sequence of enhanced features $\{\vect{z}_1, \vect{z}_2, ...\vect{z}_T\}$ with their corresponding uncertainty measures given by $-1 \times \log\,\{\vect{L}_1, \vect{L}_2, ...\vect{L}_T\}$.



\subsection{Posterior inference using frame-wise uncertainty}

We assume that a \emph{linear Gaussian model} is responsible for generating the representations $\vect{z}_t$, as follows:
\begin{equation}
\begin{split}
    \text{Generative model:}\quad \vect{z}_t &= \vect{h} + \vect{\epsilon}_t \\
    \text{Latent variable:}\quad \vect{h} &\sim \mathcal{N}\left(\vect{\mu}_{\rm{p}}, \vect{L}\inv_{\rm{p}}\right) \\
    \text{Uncertainty:}\quad \vect{\epsilon}_t &\sim \mathcal{N}\left(\vect{0}, \vect{L}\inv_t\right)
\end{split}
\label{eq:linearmodel}
\end{equation}
Here, $\vect{h}$ is the latent variable assigned to the entire sequence, and $\vect{\epsilon}_t$ is a random variable caters for the uncertainty covariance measure for each frame. We also impose a Gaussian prior on the variable $\vect{h}$ with the prior mean vector $\vect{\mu}_{\rm{p}}$ and covariance matrix $\vect{L}\inv_{\rm{p}}$. 

Given the input sequence and uncertainty estimate, it can be shown that the posterior distribution of $\vect{h}$ is also Gaussian: 
\begin{equation}
p\left( {{\bf{h}}|{{\bf{z }}_1}, \ldots ,{{\bf{z}}_T},{{\bf{L}}\inv_1},...,{{\bf{L}}\inv_T}} \right) = {\cal N}\left( {{\bf{h}}|{\phi _{\rm{s}}},{{\bf{L}}\inv_{\rm{s}}}} \right)
\end{equation}
with its posterior mean vector and precision matrix computed as
\begin{equation}
{\phi_{\rm{s}}} = {{\bf{L}}\inv_{\rm{s}}}\left[ {\sum\limits_{t = 1}^T {{\bf{L}}_t{{\bf{z }}_t}}  + {\bf{L}}_{\rm{p}}{{\bf{\mu }}_{\rm{p}}}} \right] = \sum\limits_{t = 0}^T {{\bf{A}}_t{{\bf{z }}_t}} 
\label{eq:postmean}
\end{equation}
and
\begin{equation}
{{\bf{L}}_{\rm{s}}} = {{\sum\limits_{t = 1}^T {{\bf{L}}_t}  + {\bf{L}}_{\rm{p}}}} = { {\sum\limits_{t = 0}^T {{\bf{L}}_t}}}
\label{eq:postcov}
\end{equation}
respectively. For sake of clarity, we have assigned the index $t=0$ to the prior such that ${{\bf{z }}_0} = {{\bf{\mu}}_{\rm{p}}}$, ${{\bf{L}}_0} = {{\bf{L}}_{\rm{p}}}$, and therefore
\begin{equation}
{{\bf{A }}_t} = {{{\bf{L}}\inv_{\rm{s}}} {{\bf{L}}_t} } \quad \text{for} \quad t = 0,1,...,T 
\label{eq:postgain}
\end{equation}
Notice that the posterior mean in \eqref{eq:postmean} is the weighted average of latent feature vectors ${{\bf{z}}_t}$ and prior ${{\bf{\mu}}_{\rm{p}}}$, each weighted with a gain factor ${{\bf{A }}_t}$ determined by the uncertainty measure ${{\bf{L}}\inv_t}$ and the posterior covariance ${{\bf{L}}\inv_{\rm{s}}}$. 

The posterior inference entails a temporal aggregation operation since we use one latent variable $\vect{h}$ for the entire sequence.  A similar operation was used in the classical i-vector inference. The difference lies in how the uncertainty measures are derived. In the i-vector paradigm, the uncertainty covariance matrix is drawn from a finite set (i.e., the covariance matrices of the UBM). Here, the uncertainty is predicted on the fly using a neural network. We refer to this as the \emph{heteroscedastic aleatoric} uncertainty.

\section{Xi-vector embedding with uncertainty}
\begin{figure}[t]
     \centering
     \includegraphics[width=0.40\textwidth]{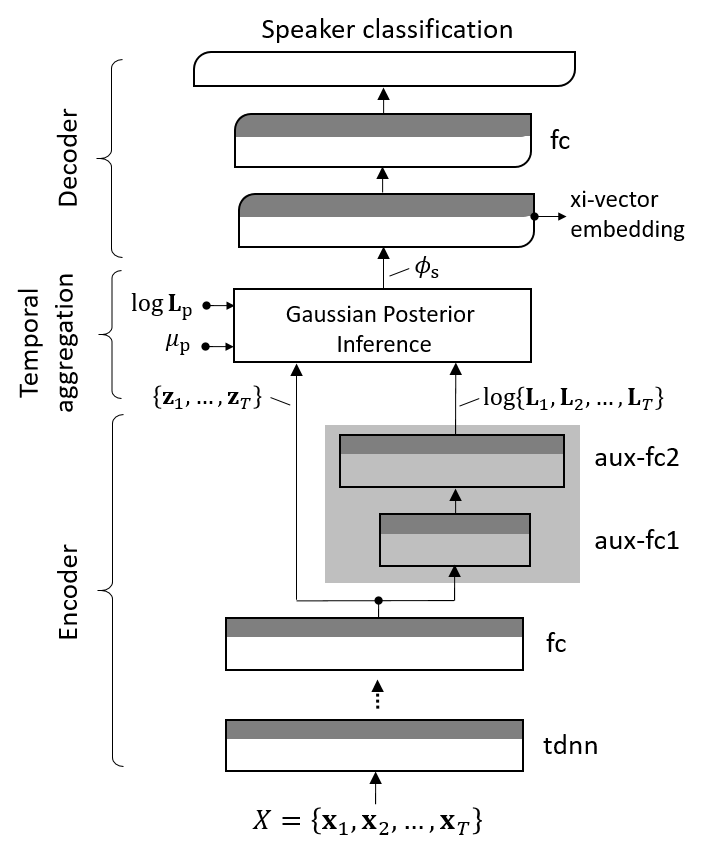}
     \caption{An exemplary xi-vector embedding neural network implemented with multiple TDNN and fully-connected (fc) layers. The round corner indicates layers that process utterance-level representations at the decoder. Shaded bars indicate non-linear activation functions. All layers use ReLU except for aux-fc2 that uses a \emph{softplus} followed by $2 \times log$ operation for log-precision estimation and a \emph{softmax} at the output.} 
     \label{fig:neural-net}
\end{figure}

We aim to incorporate frame uncertainty measures into deep speaker representation learning. This is achieved by inserting the generative linear Gaussian model into the speaker embedding neural network. Our proposal is based on the observation that Gaussian posterior inference is essentially a mapping operation where sequences of point estimates, ${\bf{z}}_t$, and uncertainty measures, ${\bf{L}}\inv_{t}$, are mapped to a fixed-length posterior mean vector. Figure \ref{fig:neural-net} shows an exemplary neural network that implements the model. Compared to the conventional x-vector, three new components are:
\begin{itemize}
    \item Frame uncertainty is inferred at the output of the frame encoder in addition to a point estimate,
    \item Gaussian posterior inference is used for temporal aggregation, and
    \item Posterior mean vector is used as input to the decoder replacing the first and second-order moments in conventional x-vector.
\end{itemize}
The posterior inference entails a temporal aggregation operation replacing the simplistic statistical pooling used in the conventional x-vector. The second-order moment is no longer required for utterance classification layers since frame uncertainty has been accounted for. As in the x-vector framework, an embedding representation is obtained by taking the pre-activation output of the first hidden layer of the decoder network.

Algorithm 1 shows the minibatch stochastic gradient descent training. In lines $6$ and $7$, the encoder produces a point estimate and the log-precision for each input frame, respectively. In lines $8$ and $9$, the algorithm estimates the gain factor using \eqref{eq:postcov} and \eqref{eq:postgain}. Since we assume that the precision matrix is diagonal and the encoder predicts directly the log-precision, the $i$-th diagonal element of the gain factor is computed as: 
\begin{equation}
    \vect{A}_t[i] = \frac{\vect{L}_t[i]}{\sum_{t'=0}^T\vect{L}_{t'}[i]} = \frac{e^{log\,\vect{L}_t[i]}}{\sum_{t'=0}^Te^{log\,\vect{L}_{t'}[i]}}
    \label{eq:softmax}
\end{equation}
for $i=1,\ldots,D$, where $D$ is the dimension of the pooling layer. Line $9$ implements Eq. \eqref{eq:softmax} with a $softmax$ function across the temporal index $t$. The entire neural network is trained by minimizing the cross-entropy loss.

Note that the prior mean and log-precision are initialized to zeros. The prior parameters are updated simultaneously with other network parameters via a posterior inference step followed by gradient back-propagation, which combines the generative nature of the i-vector with the benefit of speaker discrimination of the x-vector. Hence, we refer to the speaker embeddings as the xi-vectors -- pronounced as the /zai/ vector.


\begin{algorithm}[t]
\SetAlgoLined 
\nl Initialize: ${\bf{\mu}_{\rm{p}}} \leftarrow {\bf{0}}$, $\log\,{\bf{L}_{\rm{p}}} \leftarrow {\bf{0}}$\;
\nl \While{stopping criterion not met}{
\nl  Sample a mini-batch of $M$ examples ${(X_s,y_s)}_{s=1}^M$\;
\nl  ${\bf{g}} \leftarrow {\bf{0}}$\;
\nl \For{$s=1$ \KwTo $M$}{
\nl    $\{\vect{z}_1, \vect{z}_2, ...\vect{z}_T\} \leftarrow f_\mathrm{enc}(X_s)$\;
\nl    $\log\,\{\vect{L}_1, \vect{L}_2, ...\vect{L}_T\} \leftarrow g_\mathrm{enc}(X_s)$\;
\nl    $\log\,\vect{L}[i] \leftarrow log\,\{\vect{L}_0[i], \vect{L}_1[i],...,\vect{L}_T[i]\}$\;
\nl    $\vect{A}[i] \leftarrow softmax(\log \; \vect{L}[i]), \; i=1,2,...,D$\;
\nl    $\phi_s \leftarrow \sum_{t = 0}^T {{\bf{A}}_t{{\bf{z }}_t}}$\;
\nl    $\hat{y}_s \leftarrow f_\mathrm{dec}(\phi_s)$\;
\nl    Compute loss $L\left(\hat{y}_s,y_s\right)$\;
\nl   ${\bf{g}} \leftarrow {\bf{g}} + \frac{1}{M}\nabla L$\;
 }
\nl $\theta \leftarrow \theta + \eta{\bf{g}}$
 }
\caption{Stochastic gradient descent training of the xi-vector neural network. A training example consists of a speech segment $X_s$ of $T$ frames and a speaker label $y_s$. The set $\theta$ includes network parameters, the prior mean, and log-precision.}
\end{algorithm}

\section{Experiments}
We present two sets of experiments. The first set was conducted on the Speakers in the Wild (SITW) \emph{dev} and \emph{eval} sets \cite{McLaren2016}. The training data was drawn from the VoxCeleb-1 and VoxCeleb-2 corpora \cite{Nagrani2020}. The second set of experiments was conducted on the NIST SRE'18 and SRE'19 \emph{eval} sets. The training set consists primarily of English speech corpora, which encompasses Switchboard, Fisher, and the MIXER corpora used in SREs 04 -- 06, 08, and 10. Since the SRE eval sets consist of enrollment and test segments in Tunisian Arabic, domain adaptation was performed on the PLDA using the unlabeled subsets provided for the evaluation. We used 40 and 23-dimensional MFCCs with 10ms frameshift for the first and second sets of experiments, respectively. Mean-normalization over a sliding window of 3s and energy-based VAD were then applied. Data augmentation \cite{tomko2017} was performed on the training sets using the MUSAN dataset \cite{musan}. As in most state-of-the-art implementations, speaker embeddings were reduced to 200 dimensions via LDA projection before PLDA.    

\addtolength{\tabcolsep}{-2pt}  
\begin{table}[t]
\caption{Performance comparison of xi-vector embedding with x-vector on the SITW dev and eval sets.}
\vspace{-2ex}
\centerline{
\begin{tabular}{l c c c c}
\toprule
& \multicolumn{2}{c}{SITW-Dev} & \multicolumn{2}{c}{SITW-Eval} \\
\cmidrule(lr){2-3} \cmidrule(lr){4-5}
& EER (\%) & MinDCF & EER (\%) & MinDCF \\
\midrule
x-vector ($\mu,\sigma$)     & 2.38   & 0.251   & 2.65   & 0.278 \\
xi-vector ($\phi,\sigma$)   & 1.72   & 0.216   & 2.30   & 0.239 \\
x-vector ($\mu$)            & 2.93   & 0.333   & 3.23   & 0.362 \\
xi-vector ($\phi$)          & 1.96   & 0.214   & 2.19   & 0.248 \\ 
\bottomrule
\end{tabular}}
\label{table:sitw} 
\end{table}
\addtolength{\tabcolsep}{2pt}  

\addtolength{\tabcolsep}{-2pt}  
\begin{table}[t]
\caption{Ablation analysis of xi-vector on SITW dev and eval sets.}
\vspace{-2ex}
\centerline{
\begin{tabular}{l c c c c}
\toprule
& \multicolumn{2}{c}{SITW-Dev} & \multicolumn{2}{c}{SITW-Eval} \\
\cmidrule(lr){2-3} \cmidrule(lr){4-5}
& EER (\%) & MinDCF & EER (\%) & MinDCF \\
\midrule
No-Prior ($\phi$)            & 1.96   & 0.235   & 2.43   & 0.268 \\
Isotropic ($\phi$)           & 2.52   & 0.273   & 3.03   & 0.325 \\
No-Prior ($\phi,\sigma$)     & 2.10   & 0.240   & 2.35   & 0.254 \\
Isotropic ($\phi,\sigma$)    & 2.28   & 0.244   & 2.54   & 0.264 \\
\bottomrule
\end{tabular}}
\label{table:ablation} 
\end{table}
\addtolength{\tabcolsep}{2pt}

We first compare the performance of the proposed xi-vector to the conventional x-vector baseline. We used a TDNN-5 base architecture \cite{snyder2018vector,okabe2018attentive} for both x-vector and xi-vector extraction, with a pooling layer of $D=1500$ dimensions for all experiments. For the xi-vectors, frame precision matrices were assumed to be diagonal and estimated with a two-layer feed-forward neural network ($1500 \times 256 \rightarrow 256 \times 1500$). The size of the hidden layer was set to $256$ so that the number of parameters of the auxiliary neural network amounts to $(1500 \times 256) \times 2$, which is the same as the number of weights required to map the standard deviation of the pooled statistics to form the x-vector, i.e., $1500 \times 512$. Note that the standard deviation is no longer required in the xi-vector. 

Table \ref{table:sitw} shows the performance comparison in terms of the EER and MinDCF ($P_{target} = 0.01$). Comparing the baseline x-vector ($\mu,\sigma$) to the xi-vector ($\phi$) in the first and last rows of Table \ref{table:sitw}, respectively, the results show that the proposed xi-vector embedding gives consistent improvement on both EER and MinDCF. The performance gain amounts to $17.5\%$ in EER and $10.9\%$ in MinDCF on the SITW-Eval set. Similar performance gain could be observed on SITW-Dev set. The results in the second and third rows of Table \ref{table:sitw} show that the use of standard deviation is not required in the xi-vector but essential for x-vector embeddings. This proves the case that frame uncertainty estimate gives a better account for variability in the inputs. Note that we used the gain factor in \eqref{eq:softmax} to compute the weighted standard deviation for the xi-vector results in the second last row of Table \ref{table:sitw}.      

 
Table \ref{table:ablation} show the results of ablation analysis. We removed the prior (denoted as \emph{No-Prior}) and further replaced the diagonal variance estimate with a spherical covariance estimate (denoted as \emph{No-Prior + Isotropic}). Comparing the results in the first two rows to that of the xi-vector ($\phi$) in Table \ref{table:sitw}, we observe consistent degradation across all decision points except the EER on the dev set. On the eval set, the EER degradation amounts to $11.2\%$ with the prior removed and further to $38.7\%$ when the isotropic variance assumption was imposed. The results in the last two rows of Table \ref{table:ablation} corresponds to the xi-vector ($\phi, \sigma$) in Table \ref{table:sitw}. Similar degradation could be observed with the prior removed and the isotropic assumption, though the degradation is considerably smaller. Comparing the last two row of Table \ref{table:ablation} to the results with x-vector ($\phi, \sigma$) in Table \ref{table:sitw}, it is clear that inclusion of frame uncertainty estimate helps. Above all, these results show that a proper frame uncertainty estimation and the use of Gaussian posterior inference formulation for temporal aggregation outperform simplistic statistical pooling which depends on the empirical mean and standard deviation estimate.                

\addtolength{\tabcolsep}{-2pt}  
\begin{table}[t]
\caption{Performance comparison of xi-vector extractor with x-vector.}
\vspace{-2ex}
\centerline{
\begin{tabular}{l c c c c}
\toprule
& \multicolumn{2}{c}{SRE'18-Eval} & \multicolumn{2}{c}{SRE'19-Eval} \\
\cmidrule(lr){2-3} \cmidrule(lr){4-5}
& EER (\%) & MinDCF & EER (\%) & MinDCF \\
\midrule
x-vector ($\mu,\sigma$)     & 7.53   & 0.503   & 6.97   & 0.510 \\
xi-vector ($\phi$)          & 7.13   & 0.483   & 6.34   & 0.478 \\ 
\bottomrule
\end{tabular}}
\label{table:sre} 
\end{table}
\addtolength{\tabcolsep}{2pt}

The second set of experiments was conducted on the SRE'18 and SRE'19 eval sets. Table \ref{table:sre} shows the performance comparison in terms of EER and MinDCF ($P_{target} = 0.01$). We observe consistent performance improvement with xi-vector on both test sets. The performance gain on the SRE'18 is $5.3\%$ in EER and $4.0\%$ in MinDCF. The performance gain is slightly higher on SRE'19 which amounts to  $9.0\%$ and $6.3\%$ in terms of EER and MinDCF, respectively. These results confirm the effectiveness of the proposed xi-vector on the narrowband SRE and wideband SITW test sets.


\section{Conclusion}
From i-vector to x-vector, and xi-vector, the central theme is speaker representation learning -- to find fixed-length representations that allow easy comparison between speakers with simple geometric operations. What sets apart the proposed xi-vector from its predecessor is the use of generative modeling in a supervised training framework. One essential element of a generative model is its ability to handle uncertainty. Discriminative embeddings, like x-vectors, do not take the uncertainty (distribution) of features into consideration. We propose to characterize the uncertainty of input sequences with frame-wise uncertainty estimates, which are then used with the point estimates to derive speaker embedding vectors. This is accomplished with a linear Gaussian model trained as part of the embedding neural network.  Given the capability to take into account frame-wise uncertainty, the proposed xi-vector embeddings exhibit improve robustness to perturbation and therefore better performance.

\section{Acknowledgements}
The authors would like to thank Mr. Hitoshi Yamamoto and Dr. Shinya Miyakawa at NEC Biometrics Research Laboratories for giving the opportunity to begin and complete this work.

\bibliography{ref}

\begin{thebibliography}{10}

\bibitem{lee2020a}
K.~A. Lee, O.~Sadjadi, H.~Li, and D.~Reynolds, ``Two decades into speaker
  recognition evaluation - are we there yet?,'' {\em Computer Speech \&
  Language}, vol.~61, p.~101058, 2020.

\bibitem{i4u2019}
K.~A. Lee, V.~Hautamaki, T.~Kinnunen, H.~Yamamoto, K.~Okabe, V.~Vestman,
  J.~Huang, G.~Ding, H.~Sun, A.~Larcher, R.~K. Das, H.~Li, M.~Rouvier, P.-M.
  Bousquet, W.~Rao, Q.~Wang, C.~Zhang, F.~Bahmaninezhad, H.~Delgado, and
  M.~Todisco, ``{I4U} submission to {NIST SRE} 2018: Leveraging from a decade
  of shared experiences,'' in {\em Proc. Interspeech}, pp.~1497--1501, 2019.

\bibitem{Matejka2020}
P.~Matejka, O.~Plchot, O.~Glembek, L.~Burget, J.~Rohdin, H.~Zeinali, L.~Mosner,
  A.~Silnova, O.~Novotny, M.~Diez, and J.~Černocky, ``13 years of speaker
  recognition research at {BUT}, with longitudinal analysis of nist sre,'' {\em
  Computer Speech \& Language}, vol.~63, p.~101035, 2020.

\bibitem{Villalba2020}
J.~Villalba, N.~Chen, D.~Snyder, D.~Garcia-Romero, A.~McCree, G.~Sell,
  J.~Borgstrom, L.~P. García-Perera, F.~Richardson, R.~Dehak, P.~A.
  Torres-Carrasquillo, and N.~Dehak, ``State-of-the-art speaker recognition
  with neural network embeddings in {NIST SRE18} and {Speakers} in the {Wild}
  evaluations,'' {\em Computer Speech \& Language}, vol.~60, p.~101026, 2020.

\bibitem{Nagrani2020}
A.~Nagrani, J.~S. Chung, W.~Xie, and A.~Zisserman, ``Voxceleb: Large-scale
  speaker verification in the wild,'' {\em Computer Speech \& Language},
  vol.~60, p.~101027, 2020.

\bibitem{lee2020nec}
K.~A. Lee, H.~Yamamoto, K.~Okabe, Q.~Wang, L.~Guo, T.~Koshinaka, J.~Zhang, and
  K.~Shinoda, ``{NEC-TT} system for mixed-bandwidth and multi-domain speaker
  recognition,'' {\em Computer Speech \& Language}, vol.~61, p.~101033, 2020.

\bibitem{snyder2018vector}
D.~Snyder, D.~Garcia-Romero, G.~Sell, D.~Povey, and S.~Khudanpur, ``X-vectors:
  Robust {DNN} embeddings for speaker recognition,'' in {\em Proc. ICASSP},
  pp.~5329--5333, 2018.

\bibitem{ioffe2006plda}
S.~Ioffe, ``Probabilistic linear discriminant analysis,'' in {\em roceedings of
  the 9th European Conference on Computer Vision}, 2006.

\bibitem{Princepaper}
S.~J.~D. Prince and J.~H. Elder, ``Probabilistic linear discriminant analysis
  for inferences about identity,'' in {\em Proc. ICCV}, pp.~1--8, 2007.

\bibitem{TAKAHASHI2014}
K.~Takahashi and T.~Murakami, ``A measure of information gained through
  biometric systems,'' {\em Image and Vision Computing}, vol.~32, no.~12,
  pp.~1194--1203, 2014.

\bibitem{Reynolds00}
D.~A. Reynolds, T.~F. Quatieri, and R.~B. Dunn, ``Speaker verification using
  adapted gaussian mixture models,'' {\em Digital Signal Processing},
  pp.~19--41, 2000.

\bibitem{kenny2008study}
P.~Kenny, P.~Ouellet, N.~Dehak, V.~Gupta, and P.~Dumouchel, ``A study of
  interspeaker variability in speaker verification,'' {\em IEEE Transactions on
  Audio, Speech, and Language Processing}, vol.~16, no.~5, pp.~980--988, 2008.

\bibitem{Dehak10frontend}
N.~Dehak, P.~Kenny, R.~Dehak, P.~Dumouchel, and P.~Ouellet, ``Front end factor
  analysis for speaker verification,'' {\em IEEE Transactions on Audio, Speech
  and Language Processing}, vol.~19, no.~4, pp.~788--798, 2010.

\bibitem{okabe2018attentive}
K.~Okabe, T.~Koshinaka, and K.~Shinoda, ``Attentive statistics pooling for deep
  speaker embedding,'' in {\em Proc. Interspeech}, pp.~2252--2256, 2018.

\bibitem{qwang2018}
Q.~Wang, K.~Okabe, K.~A. Lee, H.~Yamamoto, and T.~Koshinaka, ``Attention
  mechanism in speaker recognition: What does it learn in deep speaker
  embedding?,'' in {\em Proc. IEEE SLT Workshop}, pp.~1052--1059, 2018.

\bibitem{campbell2006}
W.~M. Campbell, D.~E. Sturim, and D.~A. Reynolds, ``Support vector machines
  using gmm supervectors for speaker verification,'' {\em IEEE Signal
  Processing Letters}, vol.~13, no.~5, pp.~308--311, 2006.

\bibitem{lee2011}
K.~A. Lee, C.~You, H.~Li, T.~Kinnunen, and K.~C. Sim, ``Using discrete
  probabilities with bhattacharyya measure for svm-based speaker
  verification,'' {\em IEEE Transactions on Audio, Speech, and Language
  Processing}, vol.~19, no.~4, pp.~61--870, 2011.

\bibitem{Desplanques2020}
B.~Desplanques, J.~Thienpondt, and K.~Demuynck, ``{ECAPA-TDNN: Emphasized
  channel attention, propagation and aggregation in TDNN based speaker
  verification},'' in {\em Proc. Interspeech 2020}, pp.~3830--3834, 2020.

\bibitem{Waibel89}
A.~{Waibel}, T.~{Hanazawa}, G.~{Hinton}, K.~{Shikano}, and K.~J. {Lang},
  ``Phoneme recognition using time-delay neural networks,'' {\em IEEE
  Transactions on Acoustics, Speech, and Signal Processing}, vol.~37, no.~3,
  pp.~328--339, 1989.

\bibitem{tdnn}
V.~Peddinti, D.~Povey, and S.~Khudanpur, ``A time delay neural network
  architecture for efficient modeling of long temporal contexts,'' in {\em
  Interspeech 2015}, pp.~3214--3218, 2015.

\bibitem{tcn}
S.~Bai, J.~Z. Kolter, and V.~Koltun, ``An empirical evaluation of generic
  convolutional and recurrent networks for sequence modeling,'' {\em CoRR},
  vol.~abs/1803.01271, 2018.

\bibitem{zeinali2019}
H.~Zeinali, S.~Wang, A.~Silnova, P.~Matějka, and O.~Plchot, ``{BUT} system
  description to voxceleb speaker recognition challenge 2019,'' 2019.

\bibitem{Chung2018}
J.~S. Chung, A.~Nagrani, and A.~Zisserman, ``Voxceleb2: Deep speaker
  recognition,'' in {\em Proc. Interspeech 2018}, pp.~1086--1090, 2018.

\bibitem{Kendall2017}
A.~Kendall and Y.~Gal, ``{What Uncertainties Do We Need in Bayesian Deep
  Learning for Computer Vision?},'' in {\em Proceedings of the 31st
  International Conference on Neural Information Processing Systems},
  NIPS’17, p.~5580–5590, 2017.

\bibitem{McLaren2016}
M.~McLaren, L.~Ferrer, D.~Castan, and A.~Lawson, ``The {Speakers in the Wild
  (SITW)} speaker recognition database,'' in {\em Proc. Interspeech},
  pp.~818--822, 2016.

\bibitem{tomko2017}
T.~Ko, V.~Peddinti, M.~S. Daniel~Povey, and S.~Khudanpur, ``A study on data
  augmentation of reverberant speech for robust speech recognition,'' in {\em
  Proc. IEEE ICASSP}, pp.~5220--5224, 2017.

\bibitem{musan}
D.~Snyder, G.~Chen, and D.~Povey, ``{MUSAN}: a music, speech, and noise
  corpus,'' in {\em arXiv:1510.08484}, 2015.

\end{thebibliography}
\bibliographystyle{ieeetr}

\end{document}